\documentclass{article}
\usepackage[square]{natbib}
\usepackage{graphicx}
\usepackage{epsfig}
\usepackage{amssymb}
\usepackage{multirow}

\begin{document}

\title{Networks from gene expression time series: characterization of
correlation patterns}
\author{D. Remondini$^{1,2,3}$\thanks{Corresponding author: Gastone.Castellani@unibo.it.
D. R. and N. N. contributed equally to the paper.}, N.
Neretti$^{2,4}$, J. M. Sedivy$^5$, C. Franceschi$^2$,\\ L.
Milanesi$^6$, P. Tieri$^2$, G.C. Castellani$^{1,2,3,4}$ \and
$^1$DIMORFIPA, Universit\'a di Bologna, IT \and $^2$CIG,
Universit\'a di Bologna, IT \and $^3$INFN Sezione di Bologna, IT
\and $^4$IBNS, Brown University, Providence, USA \and
$^5$Department of Molecular Biology, Cell Biology and
Biochemistry,\\ Brown University, Providence, USA \and
$^6$ITB-CNR, Milano, IT }

\maketitle{}

\begin{abstract}
We address the problem of finding large-scale functional and
structural relationships between genes, given a time series of
gene expression data, namely mRNA concentration values measured
from genetically engineered rat fibroblasts cell lines responding
to conditional cMyc proto-oncogene activation. We show how it is
possible to retrieve suitable information about molecular
mechanisms governing the cell response to conditional
perturbations. This task is complex because typical
high-throughput genomics experiments are performed with high
number of probesets ($10^3-10^4$ genes) and a limited number of
observations ($<10^2$ time points). In this paper we develop a
deepest analysis of our previous work \citep{CmycPNAS} in which we
characterized some of the main features of a gene-gene interaction
network reconstructed from temporal correlation of gene expression
time series. One first advancement is based on the comparison of
the reconstructed network with networks obtained from randomly
generated data, in order to characterize which features retrieve
real biological information, and which are instead due to the
characteristics of the network reconstruction method. The second
and perhaps more relevant advancement is the characterization of
the global change in co-expression pattern following cMyc
activation as compared to a basal unperturbed state. We propose an
analogy with a physical system in a critical state close to a
phase transition (e.g. Potts ferromagnet), since the cell responds
to the stimulus with high susceptibility, such that a single gene
activation propagates to almost the entire genome. Our result is
relative to temporal properties of gene network dynamics, and
there are experimental evidence that this can be related to
spatial properties regarding the global organization of chromatine
structure \citep{GlobChrom2006}.
\end{abstract}

%%%%%%%%%%%%%%%%%%%%%%%%%%%%%%%%%%%%%%%%%%%%%%%%%%%%%%%%%%%%%%%%%%%%%%%%%%%%%%%%%%%%%%%%%%%%%%%

%\newpage
\section{Introduction}
Complex network theory has been used to characterize topological
features of many biological systems such as metabolic pathways,
protein-protein interactions, and neural networks
\citep{JeongBarabasi,MaslovSneppen,BarabasiOltvai,Latora}. The
application of network theory to gene expression data has been not
fully investigated, particularly regarding the relationships
between genes occurring while their expression level changes in
time. Gene expression measurements  have allowed an unbiased
search at the genome level but the large amount of experimental
data that they generate (e.g. $10^5$ probesets for human
microarrays), as well as their complexity, has slowed down
reliable modelling. An emerging and powerful approach to tackle
the complexity in functional genomics experiments, is the so
called \textit{perturbation method}, that consists in perturbing
the system with external tunable stimuli and following the changes
in the gene interaction-network properties as a function of time
and perturbation magnitude \citep{CmycPNAS,Perturb2001}. An
experimental example of this strategy is the dataset obtained from
measurements in which a single, but very important gene was
conditionally switched on or off. To do this, two cell lines were
genetically engineered so that one has served as a negative
control (the gene was absent) and the other was provided of a
reconstituted  c-Myc that allowed the conditional activation by
tamoxifen \citep{Sedivy}. In this experiments, the expression
level (mRNA concentration) of about 9000 genes was observed at
different times, generating two time series (the basal-control and
the activated-perturbed) of 5 points with 3 replicates each.

One of the key points for the application of network methodologies
to genomics data is the definition of the links between elements
(nodes), namely, the gene interactions from which all network
properties are obtained. Several methods for links assessment have
been proposed for gene expression data, such as Linear Markov
Model-based methods \citep{Maritan,Dewey} or correlation-based
methods \citep{Butte,Eisen}. We define links on the basis of the
time correlation properties of gene expression measurements, as
described in \citep{CmycPNAS}.

We analyze the distributions of the most typical network
parameters, like connectivity degree, clustering coefficient,
betweenness centrality, and the degree-degree correlation feature.
The features that characterize a specific network are those that
significantly differ from a "reference", unstructured network
model: typically the Erd\"{o}s-Renyi (\textbf{ER}) random network
is considered for such scope \citep{Newman}. In our case, the step
of calculating the correlation matrix is a processing that
introduces some structures not found in the ER "null" hypothesis,
thus we will consider as a reference the network which is obtained
from the correlation matrix of randomly generated vectors of the
same size than the real dataset.

We also show that correlation properties of gene expression time
series measurements reflect very broad changes in genomic
activity. The network features observed with real data can be
explained by a model in which correlation propagates into a large
portion of the genome, as should be expected in proximity of a
phase transition when referring to the "correlation length" of
fluctuations.

We show that it is possible, using appropriate experimental
condition, the detection, as well as the modelling of global
changes in selected group of genes, such as temporal
synchronization of gene expression dynamics.

A mechanism that is emerging as a possible explanation for such
synchronization is a global change in chromatin structure elicited
by transcription factors activity whose targets are responsible
for histon modification (acetylation and methylation state)
\citep{GlobChrom2006}.

%%%%%%%%%%%%%%%%%%%%%%%%%%%%%%%%%%%%%%%%%%%%%%%%%%%%%%%%%%%%%%%%%%%%%%%%%%%%%%%%%%%%%%%%%%%

\section{Network Construction}
Our network approach \citep{CmycPNAS} aims at characterizing the
relations among the elements of a complex system (the genome and
its mutual interactions). We consider the genes as the
\textit{nodes} of the network and, given the expression profile of
gene \textit{i} in time $\overrightarrow{g}_{i}={g}_{i}(t_k),
k=1,...,N$, a \textit{link} exists between genes \textit{i} and
\textit{j} if the absolute value of the correlation coefficient
$C_{ij}$
\begin{equation}
\label{NetDef} C_{ij} =
\frac{(\overrightarrow{g}_{i}-\langle\overrightarrow{g}_{i}\rangle)\cdot
(\overrightarrow{g}_{j}-\langle\overrightarrow{g}_{j}\rangle)}{\|\overrightarrow{g}_{i}-\langle\overrightarrow{g}_{i}\rangle\|\cdot
\|\overrightarrow{g}_{j}-\langle\overrightarrow{g}_{j}\rangle\|}\\
\end{equation}
exceeds a defined threshold. Only high values of correlation are
considered, in order to reduce spurious correlations due to noise.
We considered $|C_{ij}|> 0.97$, because this is, in absolute
value, higher than those requested for the statistical
significance of the correlation coefficients ($p<0.05$). We remark
that similar results are obtained for values in the range
[0.95:0.98]. The result is an undirected topological network
specified by its adjacency matrix $A_{ij}$, from which the self
links are removed (self-correlation is not relevant).

Three datasets are considered for network reconstruction: the
expression time series of the genes that significantly responded
to cMyc activation, referred to as the T dataset (1191 genes
observed at 5 time points); the expression time series of the same
genes in "nonperturbed" state (N dataset, same size as T); a
matrix of numbers sampled from the Standardized Normal
Distribution (0 mean, unit variance) of the same size (R dataset).
This allows to compare the perturbed state to a basal cellular
state, but also to discriminate between features specific to real
biological data and other due to the network construction
procedure. Another way to remove network structure is achieved
through random rewiring of the links, in a way to preserve single
node connectivity degree (both incoming and outgoing) but removing
the specificity of the connections (referred to as MS rewiring
\citep{MaslovSneppen}). For some analyses we will consider also an
ER network of the same size (1191x1191), with an average
connectivity degree that produces a giant component, as observed
in N and T datasets.

%%%%%%%%%%%%%%%%%%%%%%%%%%%%%%%%%%%%%%%%%%%%%%%%%%%%%%%%%%%%%%%%%%%%%%%%%%%%%%%%%%%%%%%%%%%%%

\section{Modeling the Phase Transition}
In response to c-Myc perturbation, gene expression profile
exhibits a strong degree of temporal correlation. The time
correlation matrix can be put into analogy with the correlation
matrix of spin states in a disordered Potts ferromagnet
\citep{WuPotts,SPC}. In the basal state (N dataset), the spins are
poorly aligned: in terms of gene expression, gene activity is
organized in small modules and a global synchronization is not
present. In the T case, spin alignment increases dramatically:
gene activities tend to correlate (or anticorrelate) in response
to a perturbation, represented in our case by artificially induced
activation of a single gene.

In order to quantify the perturbation extent, we generate a
randomly distributed dataset of the same size of the original cMyc
dataset (2976 5-dimensional random vectors sampled from a
Standardized Gaussian Distribution with 0 mean and unit variance).
Global correlation is introduced to various degrees by multiplying
different portions of the dataset (from 0\% to 100\%) by the
matrix of the singular values extracted from the T dataset (we
define this procedure as \textit{rescaling}).

\begin{figure}[P]
  \centering
  \includegraphics[scale=2.2]{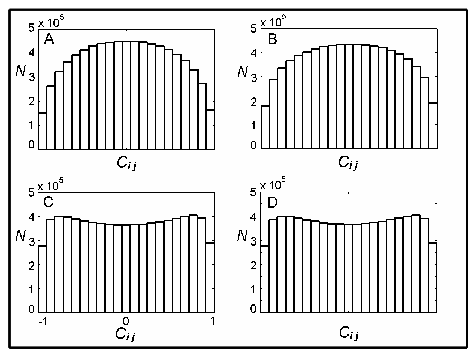}
  % bb=llx lly urx ury into (ll=lower-left, ur=upper-right)
  \caption{Histograms of the correlation coefficient distributions, obtained from real datasets and from random datasets with different degrees of correlation.
  A: N dataset. B: randomly generated dataset, 50\% correlated data, 50\% uncorrelated data.
  C: T dataset. D: randomly generated dataset, 90\% correlated data, 10\% uncorrelated data.}
  \label{Corr}
\end{figure}

A distribution of the correlation coefficients, very similar to
that observed for the T dataset, is obtained when at least 90\% of
the data are rescaled (see Fig. \ref{Corr}). No such distribution
for the correlation coefficients is obtained if rescaling is
applied up to 50-60\% of the data. Thus the cell perturbation
induced by c-Myc activation produces a synchronization of gene
activity that spans a large portion of the genome at various
degrees. Moreover, we emphasize the fact that such a large-scale
response of the genome is obtained through the perturbation
(activation) of a single element (cMyc gene). This resembles the
high susceptibility of a spin system when it is in a critical
state close to a phase transition, in which the correlation length
is of the order of magnitude of the whole system. In our case the
analogy is with a paramagnetic or antiferromagnetic system, since
the "average alignment" is very close to zero, due to an identical
amount of correlated and anticorrelated gene expression profiles
(see Fig \ref{Corr}C, or figures in \citep{CmycPNAS})

%%%%%%%%%%%%%%%%%%%%%%%%%%%%%%%%%%%%%%%%%%%%%%%%%%%%%%%%%%%%%%%%%%%%%%%%%%%%%%%%%%%%%%%%%%%%%%%%%%%%%%%%%%

\section{Network Properties}
As shown in \citep{CmycPNAS}, connectivity degree distribution
\textit{p(k)} strong\-ly reflects global changes in genome
activity. Since this feature was shown to depend on the exact time
series sequence, being disrupted by time shuffling, we argued that
it was retrieving information embedded in real data.

\begin{figure}[P]
  \centering
  \includegraphics[scale=2.2]{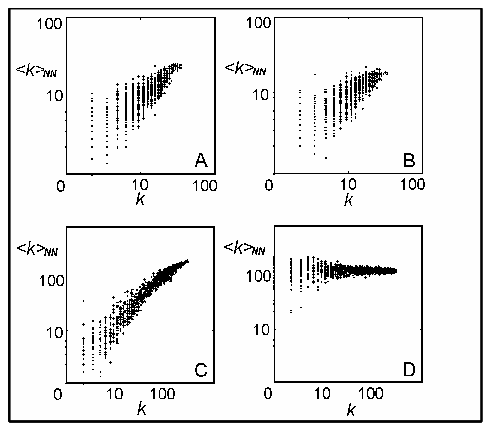}
  % bb=llx lly urx ury into (ll=lower-left, ur=upper-right)
  \caption{Degree-degree scatter plot. Logarithmic scale is
  used on both axes. X axis: connectivity degree \textit{k}; Y axis: connectivity degree averaged over the nearest neighbours for each node $\langle k \rangle _{NN}$.
  A: N network. B: R network. C: T network. D: MS rewired T network.}
  \label{KKcorr}
\end{figure}

In the present work we show that some features, on the contrary,
seem to be introduced by the processing for network construction.
For example, the clustering coefficient \textit{C} calculated for
the N and T networks is much higher than compared to a random
network of the same size ($C_{N}=0.4726$, $C_{T}=0.4975$, whereas
$C_{ER}\simeq 10^{-3}$) but the R network also has $C_{R}=0.4733$.
Another feature that is shared among the N, T, and R datasets is
the assortative property (see Fig. \ref{KKcorr}), which means that
nodes are likely to be directly connected with nodes of similar
degree. Therefore also this feature is not reflecting properties
of real data, but seems related to the thresholding procedure. We
give the following interpretation for these observations. The high
correlation relationship ($\|C\|>0.97$) can be seen as very close
to an identity relationship, and a sort of transitive property is
verified: if gene A is highly correlated to gene B, and B to C,
very likely A will be highly correlated to C. This leads to a high
density of triangles in the network (fully connected 3-node
subgraphs), that justifies the unusually high clustering
coefficient. Another effect of this transitive property is the
stratification of connectivity leading to assortativity: if a node
is highly correlated to many nodes (and thus has a high
connectivity degree \textit{k}), very likely these nodes will all
be correlated to each other (thus they will all have similar
connectivity degree). This feature is completely absent for a ER
network (data not shown) and also for MS rewiring of N and T
networks.

\begin{figure}[P]
  \centering
  \includegraphics[scale=3]{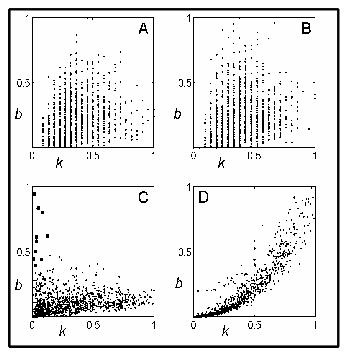} %[bb=0 0 350 320][width=2cm, height=4cm]
  % bb=llx lly urx ury into (ll=lower-left, ur=upper-right)
  \caption{Plots of betweenness centrality \textit{b} vs. connectivity degree
  \textit{k} for different realizations of the network. X axis: \textit{k}/max(\textit{k});
  Y axis: \textit{b}/max(\textit{b}). A: N network.
    B: R network. C: T network; squares: selected genes with high
    $b/k$ ratio. D: MS rewired T network.}
  \label{KvsBC}
\end{figure}

Nodes can be characterized not only by the numbers of links they
have, but also by more complex features, like betweenness
centrality \textit{b}, that characterizes the relevance for
communication between nodes in the network. Betweenness centrality
distribution for the T network is much more skewed as compared to
the N and R networks (data not shown). The joint distribution of
\textit{b} vs. \textit{k} (Fig. \ref{KvsBC}) for the T network
appears very different from the R and N cases. In particular, a
group of genes presents a high level of \textit{b} and low
\textit{k}.

We can argue that the expression profile of these genes is thus
quite different from the others (implied by the low connectivity
degree), but a lot of genes are related each other through them
(high betweenness): an hypothesis is that these are genes that
respond early to the perturbation (and thus their profile in time
is relatively unique) and guide the following gene activation
cascade (thus many genes show combinations of their time profile).
In the highest ranking genes (reported in Tab. \ref{BKtab}) with
respect to $b/k$ ratio we find genes involved in early gene
transcription events and known direct (early) targets of c-Myc
also involved in tumorigenesis.

\begin{table}[p]
  \centering
  \caption{Top 10 genes ranked by $b/k$.}\label{BKtab}
  \begin{tabular}{c|c|c}
  \textbf{B/K} & \textbf{Gene} & \textbf{Description} \\ \hline
  0.88  & Mxd3 & Max dimerization protein 3 \\
  0.70 & MGC72561 & Similar to mannosidase 2, $\alpha$ B1 \\
  0.63 & Top2a & Topoisomerase (DNA) 2 $\alpha$  \\
  0.47 & Lgals7 & Galectin-7\\
  0.39 & Cdc25b & Cell division cycle 25B\\
  0.37 & Nr5a2 & Nuclear receptor subf. 5 gr. A, m. 2\\
  0.33 & RT1-Ba & RT1 class II, locus Ba \\
  0.29 & Btnl2 & butyrophilin-like 2 (MHC class II associated) \\
  0.19 & Prkag1 & Protein kinase, AMP-activated, $\gamma$ 1 non-catalytic subunit \\
  0.11 & Ctsd & Cathepsin D
\end{tabular}
\end{table}

%%%%%%%%%%%%%%%%%%%%%%%%%%%%%%%%%%%%%%%%%%%%%%%%%%%%%%%%%%%%%%%%%%%%%%%%%%%%%%%%%%%%%%%%%%%%%%%%%%%%%%%%%%%%%%%%%%%%%%%

\section{Conclusions}
A contemporary challenge is to extract as much information as
possible from high-throughput genomic and proteomic data. These
data are usually very noisy, due to experimental techniques and
the high variability intrinsic to biological organisms. Moreover,
this problem is very often ill-posed, in the sense that the
relationships about a very high number of elements
($10^{3}-10^{4}$ genes) must be retrieved from a very small number
of samples ($<100$ experiments). It is very important thus to
consider methods able to reliably recover informations from this
kind of data.

We shown previously that it is possible to extract information
about gene-gene interaction network from a time series of gene
expression measurements, both on a global scale and with respect
to single gene roles and functions. In this paper we better
characterized the most relevant features of the networks obtained
with such method. Some features, like high clustering coefficient
or the assortativity property, appear to be dependent on the
network construction procedure (namely calculating the correlation
matrix from the data), independently of the real data
characteristics. We also show that there are relationships between
network parameters (e.g. considering betweenness centrality and
connectivity degree joint distribution) that are not found in
"null" network models. These relationships are significantly
different also when a real dataset is considered, comparing a
"basal" and a highly "perturbed" cell state. Considering the ratio
between betweenness centrality and connectivity degree of each
gene $b/k$, we obtain a ranking of the genes that seems to reflect
early response to the perturbation, and thus characterizes the
initiation of the gene activation cascade observed experimentally.

If the correlation matrix of gene expressions is seen as the
correlation of spin states of a paramagnetic/antiferromegnetic
system, the perturbation induced by a single element in the genome
(in our case cMyc proto-oncogene activation) propagates through
almost the whole system. Continuing with the analogy, such high
"susceptibility" may suggest that the genome is behaving as in the
proximity of a critical state, in which the correlation length is
of a size comparable to the entire system.

%%%%%%%%%%%%%%%%%%%%%%%%%%%%%%%%%%%%%%%%%%%%%%%%%%%%%%%%%%%%%%%%%%%%%%%%%%%%%%%%%%%%%%%%%%%%%%

\section*{Acknowledgments}
D.R. and G.C. thank Italian MURST FIRB Grant, INFN FB11 Grant. and
RFO (ex 60\%) Grant. J.M.S. acknowledges NIH grant R01 GM41690 for
support of this project

%%%%%%%%%%%%%%%%%%%%%%%%%%%%%%%%%%%%%%%%%%%%%%%%%%%%%%%%%%%%%%%%%%%%%%%%%%%%%%%%%%%%%%%%%%%%%%%%%

\end{document}